\documentstyle[preprint,prd,tighten,aps,eqsecnum,epsf,axodraw,epsf]{revtex}
\begin{document}
\preprint{\baselineskip 18pt{\vbox{\hbox{OUTP-98-11P}
\hbox{hep-th/9802119} \hbox{February 1998}}}}
\title{Induced magnetic moments in 
three-dimensional gauge theories with external 
magnetic fields}

\vspace{15mm}

\author{Nick E. Mavromatos\footnote{PPARC Advanced Fellow} and Arshad Momen }
\vspace{15mm}
\address{ University of Oxford,
Department of Physics, Theoretical Physics,\\
1 Keble Rd., OX1 3NP, U.K.}
\vspace{1mm}
\maketitle
\begin{abstract}
We study the appearance of induced  parity-violating magnetic
moment, in the presence of external magnetic fields,  
for even-number of fermion species coupled to dynamical fields in
three dimensions.  Specifically,  we use a  
$SU(2) \times U(1)$ gauge model for dynamical gauge symmetry breaking,
which has also been 
proposed recently as a field theorerical model for high-$T_c$   
superconductors.  By decomposing the fermionic  degrees of freedom
in terms of Landau levels, we show that, in the effective theory with
the lowest Landau levels, a parity-violating magnetic moment
interaction is induced by the higher Landau levels when the 
fermions are massive. The possible relevance of this
result for a recently observed phenomenon in high-$T_c$ superconductors is
also discussed. 
\end{abstract}
\vspace{5mm}

\newcommand{\be}{\begin{equation}}
\newcommand{\pr}{\paragraph{}}
\newcommand{\nn}{\nonumber}
\newcommand{\ee}{\end{equation}}
\newcommand{\bea}{\begin{eqnarray}}
\newcommand{\eea}{\end{eqnarray}}
\newcommand{\real}{{\rm l}\! {\rm R}}
\newcommand{\ra}{\rightarrow}
\newcommand{\tr}{{\rm tr}\;}
\newcommand{\al}{\alpha}
\newcommand{\del}{\Delta}
\newcommand{\Th}{\Theta}
\newcommand{\td}{\tilde{\del}}
\newcommand{\g}{\gamma}
\newcommand{\bt}{\beta}
\newcommand{\bchi}{\bbox{\chi}}
\section{Introduction}

The generation of particle masses via dynamical symmetry has been
studied in particle physics over three decades since the
pioneering work of Nambu and Jona-Lasinio \cite{nambu}. 
Apart from the realm of particle physics, this mechanism is also
responsible for generating the mass gaps in condensed-matter systems
like the BCS-type superconductors. 
Recently, it was found~\cite{klim,gusynin,leung,hys} that an  
external magnetic field
can enhance such symmetry breaking, and specifically it leads to
dynamical  chiral
symmetry breaking in QED~\cite{gusynin}.  
The dynamically generated fermion mass
then depends on the value of the external field. One can establish
this fact by examining 
the two point function fermions in the presence of
the external magnetic field\cite{schwinger}. However, due to the
presence of 
dynamical QED interactions ( apart from the external magnetic 
field), these two point functions are required to satisfy the
Schwinger-Dyson (SD) equations. 
For strong magnetic fields, the effective 
fermionic degrees of freedom are the ones in the 
lowest Landau levels (LLLs), as the energy gap
between the Landau levels (LLs) becomes very large. 
The condensate is then obtained by considering the coincidence 
limit of the two-point function for these LLL fermions, in the presence of the
external magnetic field.

In the context of 
four-dimensional QED, Hong has computed~\cite{hong} the Wilsonian effective action involving
these LLL fermions by {\em integrating out } the higher Landau levels.
The resulting  action contains four-fermi interactions apart from 
the usual minimal coupling terms. These extra interactions are responsible 
for the appearance of the non-vanishing condensate for the LLL fermions.
Though these four-fermi operators appear  
irrelevant ( in a renormalization-group sense) via
naive power counting, they 
indeed  have non-trivial anomalous-dimensions which depend on the 
magnitude of the external magnetic field. Thus, above a critical value of the 
field, they become relevant operators~\cite{hys}
and generate the chiral symmetry breaking.
The situation is quite similar to the case of three-dimensional
multiflavor four-fermi theories, where naive power counting arguments 
contradict the relevance of the interactions, which can be established 
after a large-$N$ analysis \cite{warr}.
 
The above-mentioned mass generation (at zero-temperature)  also occurs
in (2+1) dimensional QED \cite{shpagin} as well as in the 
non-Abelian theory\cite{new}. This result can be  relevant for  
high-temperature superconductors  as
effective field theories, like 
QED$_3$ ( and variants of it )\cite{NN} and 
non-Abelian gauge theory based on the group
$SU(2) \times U(1)$, can demonstrate superconductivity \cite{fm,kostas}
\footnote{ The relativistic
(Dirac) nature of the fermion fields is justified by the fact that
they  describe the excitations about the {\em
nodes} of a $d$-wave superconducting gap.} .  Indeed, there is
experimental evidence 
for the opening of a second (superconducting) gap at the nodes  of the
Fermi surfaces in certain d-wave superconductors in the presence of strong
external magnetic fields \cite{krishna}.
In \cite{kostas}, a rudimentary finite-temperature analysis was
performed and it was found that the (magnetically induced ) condensate
disappears at a critical temperature, $T_c$ which scaled with the
external magnetic field $B$ as,
\be
T_c \sim \alpha \ln \left| \frac{\sqrt{eB}}{\alpha} \right|,
\label{1.1}
\ee
with $\alpha$ being the `fine structure' constant, with dimensions of
`mass', for the $U(1)$ gauge group responsible for the generation of a
mass gap for the charge-carrying fermions, which 
is enhanced by the presence of $B$. Note that this $U(1)$ gauge theory
is not not necessarily to be identified with electromagnetism.
In (\ref{1.1}), $e$ is the electric charge ( kept four-dimensional), 
and $\frac{1}{\sqrt{eB}}$  is the `magnetic length'.

Though there is no conclusive evidence for parity and
time-reversal violation in high-temperature superconductors ( see, 
however \cite{expt} for a
recent observation for time-reversal violation ), Laughlin 
\cite{laughlin} has suggested that during the abovementioned phase transition, the order parameter develops a P and T-violating state component, induced by
the magnetic field. Therefore, it
is interesting to ask whether the effective field theories mentioned
above also lead to parity-violation in the presence of the 
external magnetic field. 
At first sight, the Vafa-Witten
theorem~\cite{vafa} prevents one from obtaining such results,
given that parity, as being 
a vector-like  symmetry,  cannot be broken in a vector-like theory, such as 
QED. Indeed, we note that 
the fermion condensates obtained in \cite{gusynin,kostas} are all
parity-conserving. This is due to the fact that 
the effective Lagrangian for
the LLL's obtained in  \cite{hong} involve parity-invariant terms only.
However, the presence of these four-fermi terms violates one of the assumptions of
the Vafa-Witten theorem which requires  the effective action to be 
quadratic in fermion fields. 
Therefore, a possibility of parity-violation  cannot be
ruled out {\it a priori} due to the presence of the fermi interactions.

A naive dimensional reduction of the 4-dimensional effective
Lagrangian  of \cite{hong} leads to a (2+1) dimensional theory with
even number of fermion flavors and it  does not lead to a parity violating 
condensate in (2+1)
dimensions as well. The flavor number needs to be even if the theory
is to describe high-$T_c$ superconducting system as  they have an  
antiferromagnetic structure. Accordingly, the system comprises of 
two sublattices and  within 
a spin-charge separation framework~\cite{Anderson}, there will be  
two species of charged fermion excitations ( called holons), one
associated with each sublattice~\cite{NN,fm}.

In this article, however, we shall show that in the presence of the 
external field,   
charged fermions in such systems also attain 
a parity-violating induced magnetic moment, in their massive 
( i.e. superconducting ) phase. This result can be established
by studying the three-point vertex
corrections depicted in Fig. \ref{fig1}. 
The internal lines in these graphs 
include fermions in higher Landau levels. 
In this work, we compute these one-loop corrections 
in  the $SU(2) \times U_S(1)$ model of \cite{farak}, which
is a toy-model for dynamical gauge symmetry breaking \cite{farak} and
has a rich phase structure \cite{fm} which can be of relevance for
superconductors.

Interestingly, graphs similar to Figure \ref{fig1} lead to 
magnetic moment for massive neutrinos \cite{mohapatra}. 
In the latter  situation, the role of the massive gauge bosons
is played by the 
massive $W^\pm$ electroweak bosons while the role of the holon 
is played by the ``massive'' neutrino.

Naively, it seems that the induction of a magnetic 
moment for a spinless fermion invalidates the initial assumption of the  
spin-charge separation in superconductors,  
given that these fermions do not carry any spin degrees of freedom. 
However, in (2+1) dimensions one can define magnetic moment without
resorting to a spin degree of freedom \cite{stern}. This is because,  
in $(2+1)$-dimensions,
the $2\times 2$ dimensional $\gamma $ matrices satisfy a $SO(2,1)$ algebra (in Minkowskian 
space times): 
\be
 [\gamma ^\mu , \gamma ^\nu ] \equiv 2 \sigma ^{\mu\nu} = 
-2i\epsilon ^{\mu\nu\lambda} \gamma _\lambda.  
\label{so}
\ee
Accordingly, the Pauli magnetic moment term can be written in terms of 
a (non-minimal) coupling to
a current, $J_\lambda = {\overline \Psi} \gamma _\lambda \Psi $,
for the Dirac fermion field $\Psi$:
\be 
{\overline \Psi} \sigma ^{\mu\nu} \Psi F_{\mu\nu} = \frac{1}{2}
\epsilon^{\lambda\mu\nu}F_{\mu\nu}J_\lambda 
\label{paulirelation}
\ee
This implies that an induced magnetic moment interaction 
for holons in planar superconductors
could still be consistent with the idea of spin-charge separation.

The paper is organized as follows. In section 2, we give a brief
review  of the $SU(2) \times U_S(1)$ model of \cite{fm}, as
well as the Dirac algebra in three dimensional spacetime with an 
even number of
fermion flavors. In section 3, we present the
induced magnetic moment calculation for the massive phase of the
model, in which the fermions and some of the non-abelian gauge
bosons acxquire (dynamically) non-trivial masses. 
In the final section, we present our conclusions and 
discuss the relevance of our results
to condensed-matter systems, in particular planar high-temperature 
superconductors.

\section{The $SU(2) \times U_S(1)$ model}

The $SU(2) \times U(1)$ model of \cite{farak} is a toy model for
dynamical electroweak gauge symmetry breaking in three dimensions,
while in the context of condensed-matter systems, the $SU(2) \times
U_S(1)$ model  of \cite{fm} is based on a gauged 
{\it particle-hole symmetry}, via an appropriate extension of the  
spin-charge separation~\cite{Anderson}. The 
holons transform as a doublet under the $SU(2)$
( particle-hole) symmetry. In this respect the model is different 
from other $SU(2) \times U(1)$ spin-charge separated theories,
which are based on either direct gauging of genuine spin rotation $SU(2)$ 
symmetries~\cite{leewen}, or non-Abelian bosonization 
techniques~\cite{marchetti,yulu}. The phase diagram  
of the model of \cite{fm}, and the associated symmetry-breaking patterns, 
are quite different from these other models, 
and will be important for our purposes here.

The three-dimensional 
continuum Lagrangian of the model is given ( in Euclidean
metric, which we use hereafter ) by ~\cite{farak,kostas}, 
\be 
{\cal L} = -\frac{1}{4}(F_{\mu\nu})^2 -\frac{1}{4}({\cal G}_{\mu\nu})^2
+ {\overline \Psi} D_\mu\gamma _\mu \Psi -m {\overline \Psi} \Psi 
\label{contmodel}
\ee
where $D_\mu = \partial _\mu -ig_1 a_\mu^S - ig_2 \sigma^a B_{a,\mu} $,
and $F_{\mu\nu}$,${\cal G}_{\mu\nu}$ are the corresponding field
strengths
for an abelian (`statistical')  $U_S(1)$ gauge field $a^S_\mu$ and 
a non-abelian (`spin') SU(2) gauge field $B^a_\mu$, respectively. 
Due to the antiferromagnetic nature of the condensed matter system
the  
fermions $\Psi$ are four-component spinors. The presence of the even
number of fermion species allows us to define chiral symmetry and
parity in three dimensions - which we discuss below.
The bare mass $m $ term is parity conserving and  
has been added by hand 
in the Lagrangian (\ref{contmodel}). In the model of \cite{fm,kostas},
this term is  generated dynamically  via 
the formation of the fermion condensate
$<{\overline \Psi} \Psi >$ by the strong 
$U_S(1)$ coupling. However, for our purposes, the details of the 
dynamical mass generation is not important and hence,  
it will be sufficient to include a bare mass term for the 
holons  representing the  
mass generated by the (strongly coupled) $U_S(1)$ interactions
in the superconducting phase.

\subsection{Dirac Algebra in $(2+1)$ dimensions}

The $\gamma _\mu $, $\mu =0,1,2$,  
matrices 
span the reducible $4 \times 4$ representation of the 
Dirac algebra in three dimensions in a fermionic theory with an 
{\it even} number of fermion flavours~\cite{app}:
\bea
&~&\gamma ^0=\left(\begin{array}{cc} i{\bf \sigma}_3 
\qquad {\bf 0} \\{\bf 0} \qquad -i{\bf \sigma}_3 \end{array} \right)
\qquad \gamma ^1 = \left(\begin{array}{cc} i{\bf \sigma}_1 
\qquad {\bf 0} \\{\bf 0} \qquad -i{\bf \sigma}_1 \end{array} \right) \nn \\
&~&\gamma ^2 = \left(\begin{array}{cc} i{\bf \sigma}_2 
\qquad {\bf 0} \\{\bf 0} \qquad -i{\bf \sigma}_2 \end{array} \right) 
\label{reduciblerep}
\eea
where 
${\bf \sigma}$ are $2 \times 2$ Pauli matrices
and the (continuum) space-time is taken to have Euclidean signature.

As well known~\cite{app} there exists two $4 \times 4 $ matrices 
which anticommute with $\gamma _\mu$,$\mu=0,1,2$: 
\be
\gamma _3 =\left(\begin{array}{cc} 0 \qquad {\bf 1} \\
{\bf 1} \qquad 0 \end{array}\right), \qquad 
\gamma _5 =i\left(\begin{array}{cc} 0 \qquad {\bf 1} \\
{\bf -1} \qquad 0 \end{array}\right)
\label{gammamatr}
\ee
where the substructures are $2 \times 2$ matrices.
These are the generators of the `chiral' symmetry for 
the massless-fermion theory 
\bea
&~&   \Psi \rightarrow exp(i\theta \gamma _3) \Psi \nn \\
&~&   \Psi \rightarrow exp(i\omega \gamma _5) \Psi 
\label{chiral}
\eea
Note that these 
transformations do not exist in the fundamental two-component 
representation
of the three-dimensional Dirac algebra, and therefore 
tha above symmetry is valid for theories 
with even fermion flavours only.

For later convenience we list several useful identities of the 
Dirac algebra, to be used in this article:
\bea 
&~& \gamma ^\mu \gamma ^\nu = -\delta ^{\mu\nu} 
- \tau _3 \epsilon ^{\mu\nu\lambda} \gamma ^\lambda \quad ; \quad  
\tau _3 \equiv i\gamma_3\gamma _5= \left(\begin{array}{cc} {\bf 1}  
\qquad {\bf 0} \\{\bf 0} \qquad {\bf -1}  \end{array} \right) \nn \\
&~& \gamma ^\mu \gamma ^\lambda \gamma ^\mu = \gamma ^\lambda \nn \\
&~&\gamma ^\mu \gamma ^0 \gamma ^i \gamma ^j \gamma ^\mu = - \delta ^{ij}\gamma ^0 - 3 \tau _3 \epsilon ^{ij} \nn \\
&~& \gamma ^\mu \gamma ^i \gamma ^j \gamma ^\mu = - 3\delta ^{ij} - 
\tau _3 \gamma ^0 \epsilon ^{ij} \nn \\ 
&~&\gamma ^\mu \gamma ^j \gamma ^i \gamma ^k \gamma ^\mu 
= - \delta ^{ij}\gamma ^k - \delta ^{ik}\gamma ^j + \delta ^{jk}\gamma ^i
\label{identities}
\eea
where repeated indices denote summation, the Greek indices
are space time indices, and the latin indices are only spatial indices.

{\it Parity} in this formalism is defined as the transformation:
\be
P:~\Psi (x^0, x^1, x^2) \rightarrow -i\gamma^3 \gamma ^1 \Psi (x^0, -x^1, x^2) \label{parity}
\ee 
and it is easy to see that a parity-invariant mass term for $\Psi$ 
amounts to masses with {\it opposite} signs between the two 
species~\cite{app}, while a parity-violating one 
corresponds to masses of equal signs. 

The set of generators 
\be
{\cal G} = \{ {\bf 1}, \gamma _3, \gamma _5, 
\Delta \equiv i\gamma_3\gamma _5 \}
\label{generators}
\ee
form~\cite{farak,fm} 
a global $U(2) \simeq SU(2) \times U_S(1)$ symmetry. 
The identity matrix ${\bf 1}$ generates the $U_S(1)$ subgroup, 
while 
the other three form the SU(2) part of the group. 
The currents corresponding to the above transformations 
are:
\be
   J_\mu^\Gamma = {\overline \Psi} \gamma _\mu \Gamma \Psi 
\qquad \Gamma =\gamma _3,\gamma _5, i\gamma _3\gamma _5 
\label{currents}
\ee
and are {\it conserved} in the {\it absence} of a fermionic {\it mass}
term. 
It can be readily verified that the corresponding charges
$Q_\Gamma \equiv \int d^2x \Psi ^\dagger \Gamma \Psi $ lead
to an $SU(2)$ algebra~\cite{farak}:
\bea
 &~& [Q_3, Q_5]=2iQ_\Delta \qquad [Q_5,Q_\Delta ]=2iQ_3 \nn \\
&~&  [Q_\Delta, Q_3]=2iQ_5 
\label{chargealgebra}
\eea
In the presence of a  mass term  an anomaly is present
\be
       \partial ^\mu J_\mu^\Gamma = 2m {\overline \Psi } \Gamma \Psi 
\label{anomaly}
\ee
while the current corresponding to the generator ${\bf 1}$ 
is {\it always } conserved, even in the presence of a fermion 
mass. 

The bilinears
\bea
&~&{\cal A}_1 \equiv {\overline \Psi}\gamma _3 \Psi,  
\qquad {\cal A}_2 \equiv {\overline \Psi}\gamma _5 \Psi,  
\qquad {\cal A}_3 \equiv {\overline \Psi}\Psi 
\nn \\
&~&B_{1\mu} \equiv {\overline \Psi}\gamma _\mu \gamma _3 \Psi,~
B_{2\mu} \equiv {\overline \Psi}\gamma _\mu \gamma _5 \Psi,~
B_{3\mu} \equiv {\overline \Psi}\gamma _\mu \Delta \Psi,~\mu=0,1,2      
\label{triplets}
\eea
transform as {\it triplets} under $SU(2)$. 
The $SU(2)$ singlets are 
\be 
{\cal A}_4 \equiv {\overline \Psi}\Delta \Psi, \qquad 
B_{4,\mu} \equiv {\overline \Psi}\gamma _\mu \Psi 
\label{singlets}
\ee
i.e. the singlets are the parity violating mass term, 
and the four-component fermion number. 

We now notice that in the case 
where 
the fermion condensate ${\cal A}_3$ is generated 
dynamically, energetics 
prohibits the generation of a parity-violating 
gauge invariant $SU(2)$ term~\cite{vafa}, and so 
a parity-conserving mass term necessarily breaks~\cite{kostas}
the $SU(2)$ group down to a $\tau_3$-$U(1)$ sector~\cite{NN}, generated
by the $\sigma_3$ Pauli matrix in two-component notation. 
 Upon coupling the system to external electromagnetic 
potentials, this phase with massive-fermions shows  
{\it superconductivity}. 
The superconductivity is strongly type II~\cite{NN,kostas} as 
the Meissner penetration depth of external magnetic 
fields turn out to be  very large,~\footnote{ Real high-temperature 
superconducting oxides are  strongly type II superconductors.
Therefore, the above theoretical model is of relevance.} and hence
the study of the response of the system to 
the external electromagnetic fields is justified.

\section{External Magnetic Fields and 
Radiatively-Induced Magnetic Moment for fermions}

The point of this section is to study the induced magnetic moment
for holons $\Psi$ in the massive phase, after coupling to 
an  external magnetic field. The magnetic moment can be computed 
by  considering  the graphs in Figure \ref{fig1}. The relevant terms 
in the effective Lagrangian, induced by the vertex correction of 
figure \ref{fig1}, will have the general form :
\be
{\cal L}_{moment} =  
\int d^3x ~{\overline \Psi}( a + b \tau_3) \sigma^{\mu \nu} \Psi
F_{\mu \nu},
\label{1.2}
\ee
where the coefficients $a$ and $b$  are to be computed. Also $\sigma^{\mu \nu }
\equiv \frac{1}{2} [ \gamma^\mu, \gamma^\nu ]$ 
but  $\gamma^\mu$ now being in 
the 4 $\times$ 4 dimensional (reducible) 
representation of the 3-dimensional Dirac
algebra  ~(due to the even number of fermion species).
$\frac{1}{2}\epsilon^{ij} F_{ij}= B$ represents the
external magnetic field , applied perpendicular to the spatial
plane. Due to the identities (\ref{identities}), the magnetic moment
interaction (\ref{1.2}) can be written as:  
\be 
{\cal L}_{moment} = - \int d^3x \int d^3x B~{\overline \Psi}\gamma ^0 
( a\tau_3 + b) \Psi
\label{pauliterm}
\ee

It can readily be seen that the term $\overline{\Psi}\gamma^0 \tau_3 \Psi$
changes sign under the parity transformation
(\ref{parity}), whereas the term $\overline{\Psi}\gamma^0 \Psi$
remains invariant under the transformation (\ref{parity}).

The computation for the coefficients $a$ and $b$  is lengthy though
straightforward. Let us now describe the basic steps of the calculation
which basically involves the vertex-function corrections depicted by
the Feynman graphs in Figure \ref{fig1}.

In momentum space, the vertex function in figure \ref{fig1} is :
\be
\Gamma^\lambda (k,q)  = - 3 (g_2)^2 \int \frac{d^3p}{(2\pi)^3} \gamma^\mu
\tilde{S}(p) \gamma^\lambda \tilde{S}(p+q ) \gamma^\nu
 {\cal D}^{(W)}_{\mu \nu} ( k-p),
\label{1.5}
\ee
where  $ {\cal D}^{(W)}_{\mu \nu} (p)$ is the propagator for the three
( hence the factor 3) $SU(2)$ `massive' gauge bosons, 
appearing in the superconducting phase of the model proposed in \cite{kostas}. For convenience, we take the 
form of the massive propagators be - 
\be
 {\cal D}^{(W)}_{\mu \nu} (p) = -\frac{\delta_{\mu \nu} }{ p^2 + M_W^2},
\label{1.6}
\ee
with $M_W$ being the mass of the gauge bosons, which is generated
dynamically in this model \cite{fm}. However, there is also a
contribution from the $U_s(1)$ gauge boson, which remain massless
in the broken-symmetry 
phase of the model of \cite{fm}. This latter contribution is
also given by an expression similar to (\ref{1.6}) but with the
massless propagator for the $U_S(1)$ gauge bosons and with $g_1^2$
replacing $3(g_2)^2$ in (\ref{1.6}). We will comment on this
contribution later. 

Using (\ref{1.6}), the vertex function (\ref{1.5}) 
reads:
\be
\Gamma ^\lambda (k,q) = -3 (g_2)^2 \int \frac{d^3p}{(2\pi )^3} 
\frac{1}{(k-p)^2 + M_W^2} {\cal N}^\lambda (p, p+q),
\label{1.6.5}
\ee
where 
\be 
{\cal N}^\lambda (p, p+q) \equiv \gamma ^\mu 
{\tilde S} (p) \gamma ^\lambda {\tilde S} (p+q) \gamma ^\mu. 
\label{nlambda}
\ee

The momentum space fermion propagator $\tilde{S}(p)$ in the presence 
of an external magnetic field \cite{schwinger} can be expanded in
terms of Landau levels\cite{chodos,gusynin,shpagin}:
\be
\tilde{S} (k) = -i e^{-\frac{{\bf k}^2}{eB} } \sum^\infty_{n=0} (-1)^n
\frac{D_n (k_0,{\bf k})}{k_0^2 + m^2 +2 eBn} ,
\label{1.3}
\ee
where 
\be
D_n (k_0, {\bf k}) \equiv \left[ ( m- k_0 \gamma_0 ) \left\{( 1- i \gamma^1 \gamma^2) 
L_n(\frac{2 {\bf k}^2}{eB})
- (1 + i \gamma^1 \gamma^2)L_{n-1}(\frac{2 {\bf k}^2}{eB}) \right\}+ 
4 \left({\bf k} \cdot
\bbox{\gamma}\right)
L^1_{n-1} (\frac{2 {\bf k}^2}{eB}) \right]
\label{1.4}
\ee
where $L_n(x)$ are the Laguerre polynomials. 
For later convenience, we use the following abbreviated form 
of the fermion propagator
\be
\tilde{S} (k) = -i e^{-\frac{{\bf k}^2}{eB} } [(m - k_0\gamma _0) \left\{
(1 -i\gamma_1 \gamma _2 ){\cal A}_1 (k) 
- (1 + i\gamma _1 \gamma _2 )
{\cal A}_2 (k) \right\} + 4 \left({\bf k} \cdot
\bbox{\gamma}\right) {\cal B} (k) 
\label{aab}
\ee
where the functions ${\cal A}_i$ and ${\cal B}$ can be read off
from the expressions (\ref{1.3},\ref{1.4}), namely
\bea
{\cal A}_1 (p) &=& \sum_{n=0}^\infty \frac{ (-1)^n  L_n ( \frac{2 {\bf
p}^2}{eB})}{p_0^2 + m^2 + 2n eB} \\
{\cal A}_2 (p) &=& \sum_{n=0}^\infty \frac{ (-1)^n  L_{n-1} ( \frac{2 {\bf
p}^2}{eB})}{p_0^2 + m^2 + 2n eB} \\
{\cal B} (p) &=& \sum_{n=0}^\infty \frac{ (-1)^n  L^1_{n-1} ( \frac{2 {\bf
p}^2}{eB})}{p_0^2 + m^2 + 2n eB} 
\label{AAB}
\eea
The vertex function, obtained after performing the Dirac algebra,
is a rather cumbersome expression and among other terms, contains 
the induced-magnetic moment term. Let us now define the quantity
\be 
{\cal K} (p) \equiv m ({\cal A}_1 (p) - {\cal A}_2 (p) ) + i\tau _3 
p_0 ({\cal A}_1 (p) + {\cal A}_2 (p) ) \equiv m {\cal F}(p) + i \tau_3
p_0 {\cal G}(p),
\label{KB}
\ee
where the quantities ${\cal A}_i (p)$, $i=1,2$, ${\cal B}(p)$ are
as defined in (\ref{aab}).

The contribution to ${\cal N}^\lambda (p, p')$ from the magnetic
moment interaction is then given by 
\be
-\tau _3 \gamma ^0 \epsilon ^{ij}\{ p'_j{\cal K}(p){\cal B}(p')-
p_j{\cal B}(p){\cal K}(p') \}e^{- \frac{({\bf p}^2 + {\bf p'}^2)}{eB}}
\equiv  -\tau _3 \gamma ^0 \epsilon^{ij} 
{\cal T}_{j}(p,q)
\label{1.7}
\ee
where $p' \equiv p+q$. Note that ${\cal T}_j(p,q)$ {\it vanishes} 
identically for $p = p'$, i.e. $q=0$. 

Therefore, the magnetic moment contribution to the vertex function is
given by 
\be
\Gamma_{moment}^i ( 0,q) =  \frac{3 (g_2)^2}{(2\pi)^3} \tau_3
\epsilon^{ij} \gamma^0 \int dp_0 \int d^2 {\bf p}
\frac{{\cal T}_j(p,q)}{p_0^2 + {\bf p}^2 + M_W^2}
\label{1.7.5}
\ee 
For our purposes of dealing with a 
a low-energy effective action, 
and a constant magnetic field, 
it will be 
sufficient to evaluate ${\cal T}_j$ for {\it small} momentum transfers
$q_j = p_j ' - p_j \ra 0$, and 
$q_0 =0$. Thus,  we retain 
only the leading  order term in the Taylor expansion 
for ${\cal T}_j$ around $q_j=0$
which is linear in $q_j$. 
In such a case of soft external photons  
only the second graph of figure \ref{fig1} contributes.
To this order, one gets :  
\be
{\cal T}_j(p,q) = -\left[q_l p_j[m\{{\cal B} \partial _l {\cal F} -
{\cal F} \partial _l {\cal B} \} + i\tau _3 p_0 \{ {\cal B}\partial _l
{\cal G}- {\cal G} \partial _l {\cal B} \} +  q_j {\cal B} \left\{
m {\cal F} + i \tau_3 p_0 {\cal G} \right \}\right]e^{-\frac{2{\bf p}^2}{eB}}. 
\label{Tj}
\ee
In the above expression ${\cal F,G, B}$ are functions of $p$ only.

Now,  the matrix $\tau_3$ appear as the combination $\tau_3
p_0$ in the function ${\cal T}_j(q,p)$. As the functions ${\cal F,G,
B}$ are even under transformation $p_0 \ra - p_0$, the terms
proportional to $\tau_3 p_0$ in (\ref{Tj}) give  zero contributions to the 
vertex function $\Gamma^i(0,q)$ after doing the $p_0$ integral. 
Thus, $b=0$ in (\ref{pauliterm}), i.e. there is no parity-conserving induced
magnetic moment. The presence of such terms would violate 
the Time-Reversal symmetry.     

Dropping the terms proportional to $\tau_3$ and using the definitions
(\ref{AAB}) in (\ref{Tj}) we get,
\bea
{\cal T}_j = -m \sum^{\infty}_{k,n=0} &&\frac{(-1)^{k+n}}{ ( p_0^2 + m^2
+ 2k eB)(p_0^2 + m^2 + 2n eB  )} \left[ -\frac{ 4 p_j ( {\bf p} \cdot
{\bf q})}{eB} \left\{ L^1_{k-1} ( \frac{2 {\bf p}^2}{eB}) L_{n-1} (
\frac{2 {\bf p}^2}{eB}) \right.\right.\nn \\ &-&\left.\left. 
L^{-1}_{k} ( \frac{2 {\bf p}^2}{eB})
L^2_{n-2} ( \frac{2 {\bf p}^2}{eB}) \right\} + q_j L^{-1}_k ( \frac{2
{\bf p}^2}{eB}) L^1_{n-1} ( \frac{2 {\bf p}^2}{eB}) \right]
e^{-\frac{2{\bf p}^2}{eB}}
\label{long}
\eea
Note that when $m$ is zero
${\cal T}_j(q,p)$ vanishes identically - even when all Landau levels
are considered. This shows that, in the normal phase
these fermions ( holons ) {\em do not carry any} induced magnetic
moment.

In the dynamical mass generation 
scenario for the fermions in the model of  \cite{fm}, the mass $M_W$ of
the $SU(2)$ gauge bosons 
(in the broken $SU(2)$ phase) is proportional 
to the fermion condensate, $u$: $M_W \sim K u $, where $u =m^2$,
and $K$ is a hopping matrix element for the fermions, 
a dimensionful constant depending on the details of the 
microscopic lattice model 
(doping concentration, Heisenberg exchange energies etc). 
The magnitude of $K$ determines the precise relation 
between $M_W$ and $m$. In the context of the  
effective (continuum) gauge field theory, 
$K$ is proportional to  
(the square of) the 
gauge couplings  of the $SU(2)$ and $U_S(1)$ sectors
(which are of equal magnitude 
in the model of \cite{fm}), and hence its magnitude depends 
crucially on the region of the phase diagram of the model 
where the analysis is performed.

To simplify our analysis in this article we shall   
assume that $M_W >> m $, which may occur for strong enough gauge 
couplings.
This approximation is qualitatively sufficient to demonstrate the
induction of a small but non-trivial magnetic moment. A more complete
analysis, including loop corrections for the gauge boson propagators,
and an extension to the case 
$M_W \sim m $ or $M_W < m$, 
will be addressed in the future.

Thus, in the approximation where $M_W$ is very large we can replace
$\frac{1}{p_0^2 + {\bf p}^2 + M_W^2}$ by $\frac{1}{M_W^2}$ in 
the gauge propagator and then the expression (\ref{1.7.5}) simpifies to 
\be 
\Gamma_{moment}^i ( 0,q) \simeq 3 \frac{(g_2)^2}{(2\pi)^3 M_W^2} \tau_3
\epsilon^{ij} \gamma^0 \int dp^0 \int d^2 {\bf p}
{\cal T}_j(p,q)
\label{amplitude}
\ee
where ${\cal T}_j(p,q)$ is given by (\ref{long}).

The spatial momentum integrals can be performed easily by the use of
the identity \cite{erdelyi}
\bea
&~&\int^\infty_o x^{\alpha + \beta} e^{-x} L^{\alpha}_m(x) L^{\beta}_n(x)
dx = \nonumber \\ 
&~&(-1)^{m+n} \frac{1}{(\alpha + m +1 ) (\beta + n + 1 )  B(\alpha + m - n +1, n + 1) 
B(m + 1, \beta + n - m +1)}
\label{identity}
\eea
where $B(p,q)$ are the Euler's beta function. 

Subsequently, only terms with $k=n$ in the double sum of
(\ref{long}) survive.  Performing the final momentum integral one gets:
\be
\Gamma_{moment}^i(0,q)  \simeq \frac{ 9 \tau _3 \epsilon^{ij}q_j 
\gamma _0 (g_2)^2 m e B}{16 \pi^2 M_W^2 } \sum^\infty_{n=1} \frac{1}{( m^2 +
2n eB)^{\frac{3}{2}}}
\label{1.8}
\ee
Note that the lowest Landau level ( i.e. $n=0$) does not appear in the
sum. Thus, the magnetic moment is induced by the higher Landau levels
only.
 
Comparing (\ref{1.8}) with the contribution for a magnetic moment term
in the vertex function  
\be 
\Gamma_{moment}^i(0,q) = \mu_B \epsilon^{ij} q_j \gamma_0 \tau_3
\label{form}
\ee
we can see that the induced magnetic moment due to the massive 
$SU(2)$ gauge boson phase, is  
\be
\mu_B|_W \simeq \frac{9(g_2)^2 m eB }{16\pi^2 M_W^2} \sum_{n=1}^\infty
\frac{1}{ ( m^2 +2n eB)^{\frac{3}{2}}}
\label{mb}
\ee
The important result is to notice that in the limit $B \rightarrow 0$ 
the interaction (\ref{1.8}) {\it vanishes}, which is a nice consistency 
check of our approach, while in the limit $eB >> m^2$, we
can ignore the $m^2$ in the denominators of the series in
(\ref{mb}). Thus, we get an induced magnetic moment,
\be
\mu_B|_W \simeq \frac{9(g_2)^2 m }{32\pi^2 M_W^2 \sqrt{2eB}} 
\zeta(\frac{3}{2})
\label{mb1}
\ee
which decreases as $\frac{1}{\sqrt{eB}}$ with increasing external magnetic
field. Thererore it vanishes in the formal limit $B \ra \infty$, 
thereby making a smooth connection with the results of 
\cite{kostas}. However, in practice the limit $B \rightarrow \infty$ 
is never reached, as is the case of high-temperature supercondutors. 
Indeed, superconductivity is destroyed 
in that case by strong 
magnetic fields higher than a critical value 
of order $20$ 
Tesla. For fields smaller than this critical value,
which is the case encountered in the experiments of \cite{krishna},
the contribution (\ref{mb1}) can be important. 

Let us now discuss the magnetic moment induced by the massless
$U_S(1)$ gauge boson interactions. Again, we shall make use of 
the tree level
propagator for the massless gauge boson, which will be 
sufficient for our purposes. 
The vertex function is then given, as
before by (\ref{1.7.5}), but with $3(g_2)^2$ replaced by $g_1^2$ and
with $M_W = 0$:
\be
\tilde{\Gamma}_{moment}^i ( 0,q) = \frac{(g_1)^2}{(2\pi)^3} \tau_3
\epsilon^{ij} \gamma^0 \int dp^0 \int d^2 {\bf p}
\frac{{\cal T}_j(p,q)}{p_0^2 + {\bf p}^2}
\label{1.8.5}
\ee 
where ${\cal T}_j$ is given, as before, by (\ref{Tj}).
The momentum integrals in the above expression cannot be computed in a
closed form. However, we can attempt to contrast this result with that
discussed for the massive boson case. Therefore, assuming that the
external magnetic field  is strong we
truncate the infinite sums over the Landau Level poles
entering in the quantities appearing in (\ref{Tj}) by considering only
the lowest and the 
first excited Landau Level pole in the propagator ( \ref{1.3}).
In this case, the function ${\cal T}_j(q,p)$ assumes the form 
\be 
{\cal T}_j = \frac{m}{p_0^2 + m^2 + 2eB} \left[\frac{4p_j ({\bf p}\cdot
{\bf q})}{eB (p_0^2 
+ m^2 + 2eB)}  - 
q_j\left\{ \frac{1}{p_0^2 + m^2} 
-\frac{ \frac{2{\bf p}^2}{eB}}{p_0^2 + m^2 + 2eB}\right\} \right],
\label{firstLL}
\ee
where, as before, we have dropped the terms proportional 
to $\tau_3$, which vanish
after performing the $p_0$ integration. Using the expression
(\ref{1.8.5}) we get the integral
\bea
\int_{-\infty}^\infty &dp_0& \frac{{\cal T}_j d^2p}{p_0^2 + {\bf p}^2} \nn \\
&=&  
-2\pi m  q_j \int dp_0 \left[
\frac{1}{(p_0^2+ m^2 + 2eB)^2} - \frac{ \Phi(\frac{2p_0^2}{eB})}{p_0^2
+ m^2 + 2eB} \left\{ \frac{\frac{2p_0^2}{eB}}{p_0^2 + m^2 + 2eB} -
\frac{1}{2 (p_0^2 + m^2)} \right\} \right] \nn \\
&=& -2\pi m q_j  \left[ \frac{\pi}{2( m^2 + 2eB)^{\frac{3}{2}}} 
- \int_{-\infty}^\infty dp_0 \frac{ \Phi(\frac{2p_0^2}{eB})}{p_0^2
+ m^2 + 2eB} \left\{ \frac{\frac{2p_0^2}{eB}}{p_0^2 + m^2 + 2eB} -
\frac{1}{2 (p_0^2 + m^2)} \right\} \right]
\label{noname}
\eea
where the function $\Phi (z)$ is defined as
\be
\Phi(z) \equiv \int_0^\infty \frac{e^{-x}dx}{x + z}
\label{phi}
\ee
and is plotted in the figure \ref{fig2}.

Notice that the function $\Phi (z)$ diverges as ${\rm ln}z$ 
in the limit $z \rightarrow 0$. In the context of (\ref{noname}) 
this limit occurs for either 
$p_0 \rightarrow 0$ or $eB \rightarrow \infty$. 
The limit $eB \rightarrow \infty$ yields a zero asymptotic result 
for the magnetic moment,  
like the case of the magnetic moment induced by
the massive gauge bosons. However, 
the 
magnetic moment due to the massless gauge bosons 
falls faster with increasing $eB$, vanishing 
in the limit $B \rightarrow \infty$ at least as 
$\frac{1}{eB}$.
On the other hand, the limit $p_0 \ra 0$, yields finite contributions 
after the $p_0$ momentum integration in (\ref{noname}).

An accurate estimate of the induced moment
cannot be made at this stage due to the approximation used in above
calculations, as we have ignored the quantum corrections in the gauge
boson propagators due to fermion loops. Nevertheless, the above 
calculation is sufficient to
demostrate our main point - the inclusion of the higher Landau levels
in the loops induces P-violating terms in the effective action as
induced magnetic moments. 

As we have argued previously in the massless ( i.e.  normal) phase, 
the fermions and the 
gauge bosons are all massless and therefore the induced moment is
absent in that phase. 
In this regard, the above phenomena can be treated as an interesting
and verifiable prediction 
for gauge models of 
high-temperature superconductors \cite{NN,kostas}.

\section{Conclusions} 

In this work we have argued how in three-dimensional gauge theories
massive fermions acquire a magnetic moment 
in the presence of (strong) external magnetic fields. The relevant
Feynman 
graphs for the induction of the magnetic moment are shown  
in figure \ref{fig1}. It is essential for the phenomenon that the
fermion degrees of freedom pertaining to the higher Landau level poles
are included as internal lines in the diagrams. Even though we have
used the model of \cite{farak,fm} to demonstrate this phenomenon,
this phenomenon seems to be model independent, the only requirement
being that the fermion is massive and coupled to gauge fields in the presence
of the external magnetic field.

The induction of the magnetic moment seems paradoxical at 
first sight, given that if one restricts oneself to the LLL case from
the beginning, there is no induced magnetic moment. 
This is similar to the coupling of the neutral pions to  photons, which is
absent in the minimal-coupling scheme, but is
induced via the anomalous fermionic loops. 

An important feature of the 
induced magnetic moment
interactions is that they 
violate parity. However, this does not necessarily imply
that the fermion condensate in the presence of this extra interaction
violate parity, as suggested recently 
by Laughlin \cite{laughlin}. One way to
check this would be to include this interaction into the Schwinger-Dyson
equations and check for a self-consistent solution 
for the dynamically-generated fermion mass gap, which violates
parity.

We have also shown that the induced magnetic moment due to the massive
gauge bosons and massless gauge bosons scale differently. Therefore, a
magnetic susceptibility measurement on the superconducting samples
would establish the relevance of the models \cite{NN,fm}. In fact, the
presence of a non-vanishing magnetic moment at very strong fields will
be a strong indication in favour of 
the existence of these massive gauge-boson excitations.

It is also interesting to note that 
anyons, which interact via Chern-Simons interactions, also acquire an
induced magnetic moment \cite{stern,szabo}. However, the induced magnetic
moment discussed here scales with the external magnetic field
differently than the induced magnetic moments for anyons: for
anyons they scale as the inverse of the applied magnetic field, while
the magnetic moments discussed here do not. However, our result for the 
induced magnetic moent due to the massless gauge bosons has been obtained 
including only the first excited Landau level.  
More accurate analyses, including 
the higher Landau levels are clearly desirable.
Given that closed analytic formulas for such a case are 
probably not possible to obtain,
Lattice analyses might be necessary \cite{lattice}. Also, it would be
interesting to see whether any connection between the two theories ( in
the presence of external magnetic fields ) exist. It is also known that
such induced magnetic moments lead to short range interaction between
charged particles in three spacetime dimensions - leading to exotic
bound states \cite{szabo}. This result might again be relevant for the
formation of a possible parity violating fermion condensate,
as suggested in \cite{laughlin}.

Finally, though our calculations are set in three
dimensional space time, nevertheless it is natural to expect the 
phenomenon of the induced magnetic moment
to occur in (3+1) dimensions as well. This could be relevant 
for the physics of chiral symmetry breaking in the Early 
Universe. 
We hope to report
on this issue in the near future.

\section*{Acknowledgements}

We are grateful to K. Farakos for a critical reading of the manuscript 
and for checking some of the calculations. 
We also acknowledge discussions with 
A. Campbell-Smith, I. Kogan, G. Koutsoumbas
and R. Szabo. 
This research was funded by PPARC (UK).

\begin{figure}
\begin{center}
\begin{picture}(350,100)(0,0)
\Line(0,50)(160,50)
\GlueArc(80,50)(50,0,180){4}{10}
\ZigZag(30,50)(80,50){2}{100}
\Photon(80,50)(80,0){6}{5.5}
\Vertex(130,50){1}
\Text(170,50)[]{$,$}
\Line(180,50)(340,50)
\GlueArc(260,50)(50,0,180){4}{10}
\ZigZag(210,50)(310,50){2}{200}
\Photon(260,50)(260,0){6}{5.5}

\end{picture}

\vspace{1cm}
\caption{Vertex Corrections leading to a 
magnetic moment interaction. The curly line represents gauge bosons
and  the wavy line indicate
an external static photon (due to the magnetic field). 
The thin straight lines denote Lowest-Landau-Level fermions, while the thick
straight lines represent fermions in higher Landau levels. 
In the case of soft external photons, $ q \rightarrow 0$, 
only the second graph contributes.}
\label{fig1}
\end{center}
\end{figure}
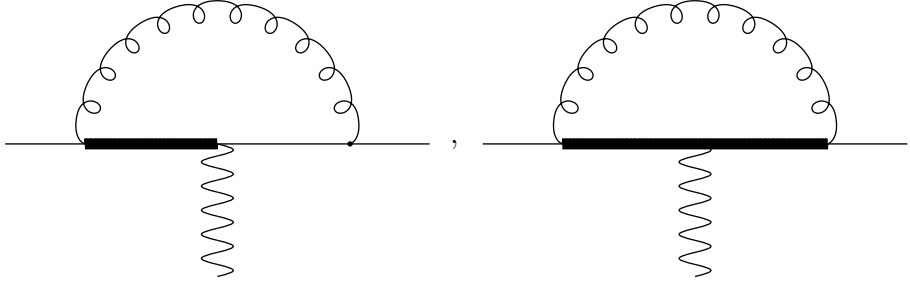
\begin{figure}
\epsfxsize=5in
\epsfysize=5in
\centerline{\epsfbox{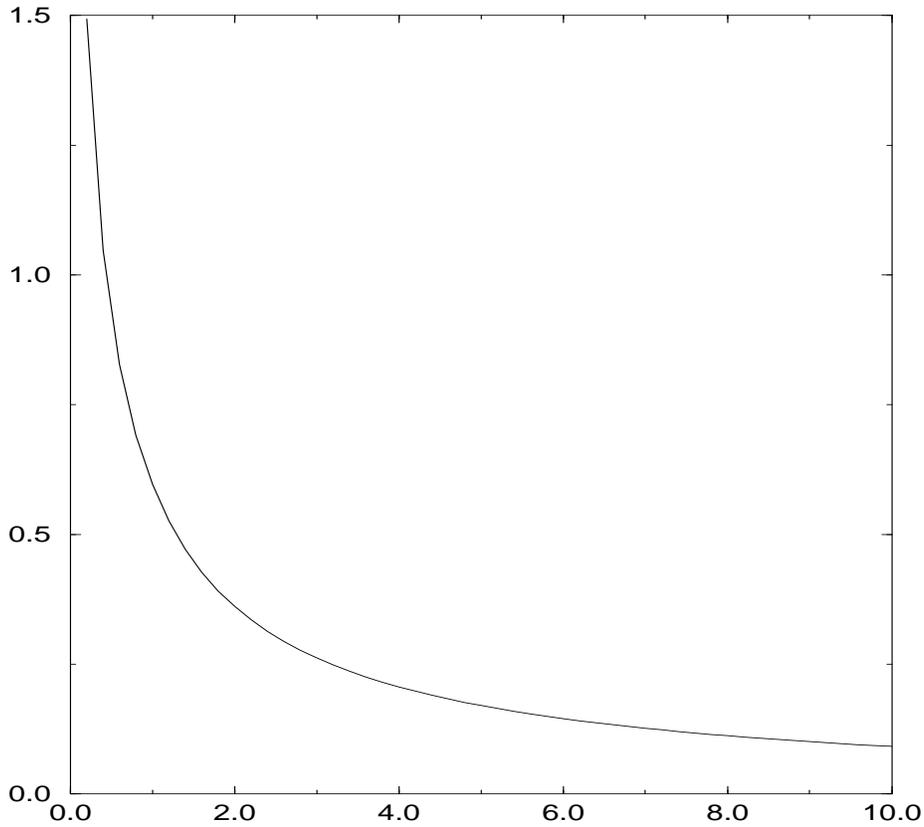}}
\caption{ The function $\Phi(z)$, defined in the text, (\ref{phi}), plotted
 vs. $z$. The function diverges as ${\rm ln}z$ as $z \ra 0$.}
\label{fig2}
\end{figure}

\end{document}